\documentclass{aa} 

\usepackage{graphicx}
\usepackage{txfonts}
\usepackage{hyperref}
\hypersetup{
    citecolor=cyan,
    colorlinks=true,
    linkcolor=red,
    filecolor=red,      
    urlcolor=magenta
    }

\begin{document}

   \title{Cluster aggregates surrounding Pismis\,5 in the Vela Molecular Ridge \thanks{The table of member candidates for eight open clusters is only available in electronic form at the CDS via anonymous ftp to cdsarc.cds.unistra.fr (130.79.128.5), or via https://cdsarc.cds.unistra.fr/cgi-bin/qcat?J/A+A/xxx/xxx.}} 

   \subtitle{}

   \author{Ming Feng Qin\inst{1,2},
   Yu Zhang\inst{1,2},
   Jinzhong Liu\inst{1,2},
   Fangfang Song\inst{1,2},
   Qingshun Hu\inst{1,2},
   Haozhi Wang\inst{1,2},
   Shuo Ma\inst{1,2}
   \and
   Guoliang L$\rm{\ddot{u}}$\inst{3}
          }

   \institute{Xinjiang Astronomical Observatory, Chinese Academy of Sciences, No.150, Science 1 Stree, Urumqi, Xinjiang 830011, People's Republic of China\\
              \email{zhy@xao.ac.cn}
         \and
         School of Astronomy and Space Science, University of Chinese Academy of Sciences, 19 Yuquan Road, Shijingshan District, Beijing 100049, People's Republic of China  
         \and 
         School of Physical Science and Technology, Xinjiang University, No.777, Huarui Street, Urumqi, Xinjiang 830017, People’s Republic of China}

   \date{Received Xxxx YY, 2022; accepted Zzzz WW, 2022}
    \authorrunning{Qin et al.}
 
  \abstract
{In the {\it Gaia} era, the precision of astrometric data is unprecedented. High-quality data make it easier to find more cluster aggregates and gather further confirmation of these 
open clusters.}   
  {We use {\it Gaia} Data Release 3 (DR3) to redetermine the open clusters surrounding Pismis\,5 
  in the Vela Molecular Ridge (VMR). We also investigate the basic properties of these
  clusters.}
  {We applied two clustering algorithms (\textsc{StarGO} and pyUPMASK) to identify the open-cluster members in five-dimensional space with $\alpha$, $\delta$, $\varpi$, $\mu_\alpha \cos\delta$, 
  and $\mu_\delta$.}
  {We identify eight open clusters surrounding Pismis\,5 in the VMR. The open cluster QZ\,1 is newly discovered.
  As a result of our investigation of the comprehensive properties of the clusters, we present one open binary cluster candidate (Alessi\,43 and Collinder\,197)
  and one triple open-cluster candidate (Pismis\,5, Pismis$\,5_{A}$, and Pismis$\,5_{B}$).}
  {We identify binary and triple open-cluster candidates as potential primordial aggregates based on their similar age, position, and motion. According to kinematic speculations, the two aggregate candidates will gradually separate, and their interiors will slowly disintegrate.}
  \keywords{ Galaxy: stellar content -- open clusters and associations: general --  methods: data analysis}

   \maketitle
%

\section{Introduction}\label{sec:intro}
Open clusters are gravitationally bound systems of stars that share a common 
heritage from the same progenitor molecular cloud, resulting in members with similar ages, distances, and metal abundances.
In particular, young open clusters contain young stellar objects in various 
evolutionary states and they are essential testbeds for understanding star formation and evolution \citep{2003ARA&A..41...57L}. 
Most open clusters are found in the Galactic disk, and their 
spatial distributions play critical roles in the study of the Galactic structure. 

The Vela Molecular Ridge $\left( {\rm VMR}; 260^{\circ} \lesssim \ell \lesssim 272^{\circ} ;|b| \lesssim 3^{\circ}\right)$ 
identified by \citet{1991A&A...247..202M} is a giant molecular cloud located in Vela and Puppis. 
\citet{1991A&A...247..202M} split the VMR into four clouds (named A, B, C, and D) according to 
local $CO$ peaks. VMR\,A, VMR\,C, and VMR\,D are situated at $700\pm200$ pc from the sun, 
while VMR\,B is located at about $2000$ pc, which means VMR\,B is not related to the other clouds. 
Therefore, our research only refers to the open clusters in the VMR\,A, VMR\,C, and VMR\,D. In the 
following, all instances of VMR refer to VMR\,A, VMR\,C and VMR\,D.

Pismis\,5 (also known as ESO\,313-SC7, $l$ = $259.355$ deg, $b$ = $0.905$ deg),
is situated in the VMR at a distance of 947.6 pc from the Sun \citep{2020A&A...633A..99C}. It is a young open cluster with an estimated age of 7.6 million years according to \citet{2021MNRAS.504..356D}. It is also a low-mass open cluster, with a flat mass function and an irregular radial density profile \citep{2009MNRAS.397.1915B}.
There are several known open clusters surrounding Pismis\,5 in the VMR,
for example, Collinder\,197, Alessi\,43, and BH\,56. 
Collinder\,197 ($l$ = $261.517$ deg, $b$ = $0.956$ deg), Alessi\,43 (also known as ASCC\,50;
$l$ = $262.546$ deg, $b$ = $1.501$ deg), and BH\,56 ($l$ = $264.470$ deg, $b$ = $1.548$ deg) are located 
at about 940 pc, 940 pc, and 903 pc, respectively, from the Sun \citep{2020A&A...633A..99C}.
Collinder\,197 and Alessi\,43 form a binary open-cluster candidate, and were recognized to do so in previous 
studies due to their identical movements and near spatial proximity \citep{2009A&A...500L..13D,2021A&A...649A..54P}.
Furthermore, \citet{2019ApJS..245...32L} use the FOF algorithm for member-star determination, and Collinder\,197, Alessi\,43, and BH\,56 are considered to be a cluster group. In this paper, we study the fundamental nature of these clusters as a step towards further understanding the VMR.

In this work, we aim to discover previously neglected clusters surrounding Pismis\,5 in the VMR through the
\textsc{StarGO}\footnote{\url{https://github.com/zyuan-astro/StarGO-OC}} \citep{2018ApJ...863...26Y}
and pyUPMASK \citep{2021A&A...650A.109P} algorithms. 
This paper is organized as follows. Section \ref{sec:member} describes the process of data processing 
and the selection of cluster members. Section \ref{sec:properties} displays the spatial and 
proper-motion distribution of clusters, as well as assessments of cluster characteristics such as 
age, stellar mass function, level of mass segregation, and initial stellar mass function. Our 
discussion and summary are presented in Section \ref{sec:discussion} and Section \ref{sec:sum}, respectively.

\section{Data and membership analysis} \label{sec:member}

Gaia data release 3 ({\it Gaia} DR3, \citealt{2016A&A...595A...1G,2022arXiv220606207G})
provides a catalog of 1.8 billion sources brighter than 21 mag in the $G$ band, including parallax
and proper motion values. 
{\it Gaia} DR3 includes 34 months of observations, a 12 month increase over
{\it Gaia} DR2, and extends two magnitudes fainter than {\it Gaia} DR2. 
Furthermore, {\it Gaia} DR3 also includes 33\,812\,183 sources with radial velocity (RV) determinations, where RVs have a median formal precision of 1.3 and 6.4 km/s at $G_{\rm RVS}$ =12 and $G_{\rm RVS}$ =14 mag, respectively. These sources form the cornerstone of studies of the three-dimensional kinematics of open clusters \citep{2022arXiv220605902K}. However, the determination of RVs is incomplete and cannot be used 
for member-star selection. Therefore, we use highly accurate 
five-dimensional astrometric data ($\alpha$, $\delta$, $\varpi$, $\mu_\alpha \cos\delta$, and $\mu_\delta$) 
provided by {\it Gaia} DR3 as the initial data with which to investigate open clusters surrounding Pismis\,5 in the VMR.

\begin{figure*}
    \centering
    \includegraphics[width=1.0\textwidth]{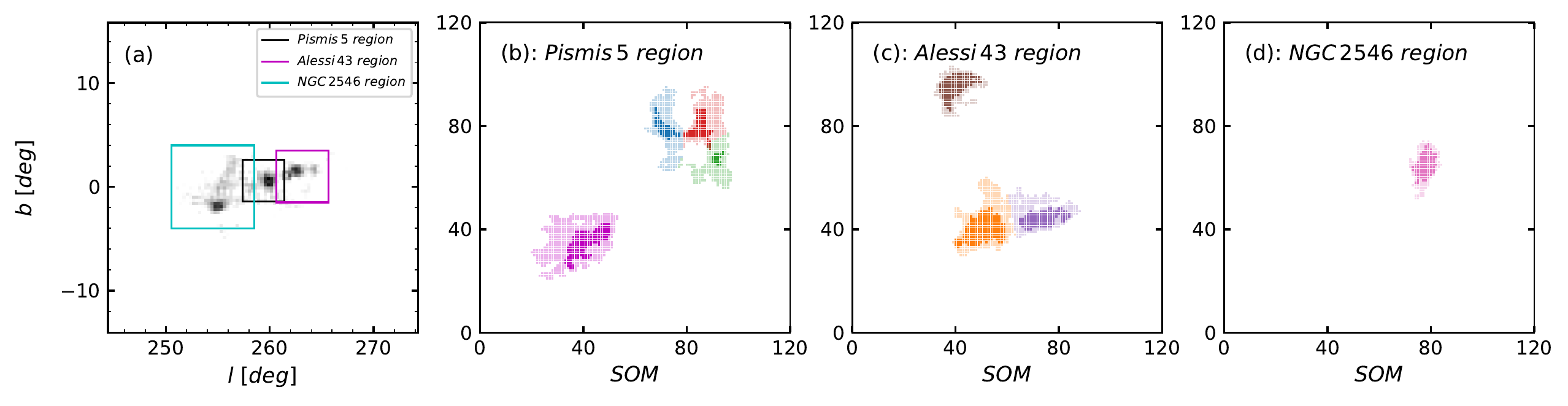}
    \caption{ Two-dimensional histogram of {\it Sample~I} in $l$-$b$ plane and the results from SOM. ~(a): 2D histogram of the spatial distribution ($l$-$b$) of     {\it Sample~I}, which only shows the bins with overdensities $> 3\sigma $. The NGC\,2546 region, 
    Pismis\,5 region, and Alessi\,43 region are each represented by a box in cyan, black, or magenta, 
    respectively. The sources in the cyan, black, and magenta boxes are referred to as
    {$\it Sample~2 $}, {$\it Sample~3 $}, and {$\it Sample~4 $}.
    ~(b), (c), and (d):  2D neural network produced by SOM, showing the neurons with a $u$ value below the 99.85th percentile of the $u$ distribution.} \label{fig:member}
    \end{figure*} 
    
\subsection{Data processing and analysis}
\label{subsec:Gaia DR3}

Firstly, the data with a radius of 250 pc centered on the Galactocentric coordinates
\footnote{The origin of the Cartesian Galactocentric coordinate system is the Galactic 
center ($l = 0^{\circ}$ and $b = 0^{\circ}$). The Galactocentric coordinates have a positive x 
direction pointing from the position of the Sun projected to the Galactic midplane to the Galactic center; the 
positive y-axis points toward $l = 90^{\circ}$ and the positive z-axis points toward b = $90^{\circ}$.}
(X, Y, Z) = ($-8474.93, -931.18, 42.54$) pc 
of Pismis\,5 were chosen in order to include as many open clusters surrounding Pismis\,5 in the 
VMR as possible. The coordinates of Pismis\,5 were 
transformed from the $\alpha$, $\delta$, and parallax values adopted from \citet{2020A&A...633A..99C} via the
Astronomical Coordinate Systems module of the \texttt{Astropy} Python 
package \citep{2013A&A...558A..33A,2018AJ....156..123A}\footnote{For the package, we adopted the values for the Galactocentric coordinate frame, namely: ICRS coordinates
($\alpha$, $\delta$) of the Galactic center $=(266.4051, -28.936175)$ deg; Galactocentric distance
of the Sun $=8.3$ kpc, and the height of the Sun above the Galactic midplane $=27.0$ pc \citep{2019ApJ...877...12T}.}.
We corrected the parallax zero-points of all the sources using the procedure described by 
\citet{2021A&A...649A...2L}. To minimize the impact of data uncertainties, we applied a series of filters to exclude artifacts and low-quality sources. We removed stars with relative uncertainties in parallax and photometry greater than $10\%$. Additionally, we discarded stars with renormalized unit weight errors (RUWEs) exceeding 1.4 because they indicate a potentially problematic astrometric solution \citep{2021A&A...649A...5F}.
The data were incomplete beyond 18 mag, because the distribution of the $G$-band of the data clearly shows a truncation at 18 mag.

High-quality data enable us to  investigate the 
open clusters in the target area  more thoroughly, despite the fact that our quality 
cut implies a certain level of incompleteness. We used a set of selection criteria to identify sources with a stellar density of greater than 50$\%$ in their proper-motion distribution in right ascension and declination. The selection criteria were $-8.89\leq {\mu}_{\alpha }cos{\delta } \leq -1.86$ mas/yr and $0.03\leq {\mu}_{\delta } \leq 8.00$ mas/yr. The reason for applying these criteria was to exclude the majority of the field stars, which would not be relevant for the more detailed data processing that follows. A total of 158,593 sources met our selection criteria and were retained as Sample 1.

Secondly, we constructed a two-dimensional (2D) histogram of the spatial distribution ($l-b$) to restrict
the {\it Sample~1} in Figure~\ref{fig:member}~(a). This diagram only represents bins with 
overdensities >3$\sigma$ in {\it Sample~1}, where $\sigma$ is the standard deviation of all bins.
The overdensity in the histogram signifies the clustering of stars. We therefore tried to include 
high-density areas for the next member star identification. 
There are three overdensities in Figure~\ref{fig:member}~(a), and the cyan box contains NGC\,2546, 
referred to here as {$\it Sample~2 $}, the black box contains Pismis\,5, referred to as {$\it Sample~3, $} and the magenta 
box contains Alessi\,43, Collinder\,197, and BH\,56, referred to as {$\it Sample~4 $}, as revealed by the cluster 
information provided by \citet{2020A&A...633A..99C}. Despite the fact that NGC\,2546 
($l$ = $254.919$ deg, $b$ = $-2.009$ deg, the blue box region in Figure~\ref{fig:member}~(a)) 
is not our target object located in the VMR, it was included in our initial sample and subsequently subjected to analysis. The sides of cyan, black, and magenta boxes in the $l-b$ plane
are 4, 2, and 3 deg in length, respectively. 
These values are a compromise between keeping more member stars and removing more 
field stars.

\subsection{Membership selection} \label{subsec:method}
There are several robust algorithms available for determining the membership of stars, including DBSCAN \citep{Ester1996}, FoF \citep{2019ApJS..245...32L}, \textsc{StarGO} \citep{2018ApJ...863...26Y}, pyUPMASK \citep{2021A&A...650A.109P}, and others. 
We combine \textsc{StarGO} and pyUPMASK to determine the open-cluster member stars.

\subsubsection{\textsc{StarGO}} \label{subsec:StarGO}

The Stars' Galactic Origin (\textsc{StarGO}) clustering method, developed by \citet{2018ApJ...863...26Y} and based on the self-organizing map (SOM). 
This algorithm will build a neural network, where each neuron is represented by a weight 
vector that is consistent with the dimension of the input data. SOM maps multidimensional input data onto a two-dimensional neural network using weight vectors. For each star we input, the weight vector of all neurons on the neural network will be changed according to the input vector. The difference in weight vectors between adjacent neurons is represented by the $u$ value; the smaller the $u$, the more similar the neurons are. Finally, the algorithm will return a series of $u$ values. In the case of uniformly distributed stellar fields, the distribution of $u$ values should conform to a normal distribution. However, if the input sample contains clusters, the distribution of $u$ values will consist of two parts: a normal distribution representing field stars and a small series of $u$ values representing clusters. Connected blocks with smaller $u$ values indicate more similar neurons, representing a group of stars that share similarities in the input data dimension. These identified groups serve as potential open cluster candidates. 
The successful application of \textsc{StarGO} in several studies ( e.g., \citealt{2019ApJ...877...12T, 2020ApJ...889...99Z, 2021ApJ...912..162P, 2022ApJ...931..156P}) further supports its reliability and effectiveness as a clustering method.

Using 1/$\varpi$ as distance, we computed Galactocentric Cartesian coordinates (X, Y, Z) for each source from {$\it Sample~2 $}, {$\it Sample~3 $}, and {$\it Sample~4 $}. We generated a
$120\times120$ network and then entered the data (X, Y, Z, $\mu_{\alpha} \cos{\delta}$, and $\mu_\delta$) for each star from {$\it Sample~2 $}, {$\it Sample~3 $}, and {$\it Sample~4 $}, respectively. 
After 400 iterations, we obtained the values of $u$ for three samples and selected the $u$ that were smaller
than the 99.85th percentile of the $u$ distribution. Our selection was equal to deleting any values greater than 
$u_{peak}$ (the peak position of the $u$ distribution), and $u$ values that were less than 
$u_{peak}$ but within 3$\sigma$ ($\sigma$ was the standard deviation of the normal distribution). As a 
result, the field star distributions were almost entirely removed. 
Neurons that passed our filter requirements were displayed as transparent patches in 
Figure~\ref{fig:member}~(b), ~(c), and ~(d).
When the threshold of $u$ was lowered, SOM displayed neurons that are more similar, which were the core structures 
inside the transparent patches. After calculating the smallest difference between 
neurons in the transparent patches and neurons in each core patch, transparent neurons were assigned to the closest core block. Their colors aligned with the colors of their core patches. Each color patch represented an open-cluster candidate. The member 
candidates of eight open clusters were obtained by associating the stars from the patches. 

\subsubsection{pyUPMASK} \label{subsec:pyUPMASK}

To reduce the contamination rate of member stars, we use pyUPMASK for member-star selection.
pyUPMASK is a modified version of Unsupervised Photometric Membership Assignment in Stellar Clusters (UPMASK, \citealt{2014A&A...561A..57K}), an open-source package written in the Python language. pyUPMASK uses clustering methods with a random initialization to identify clumps, followed by an assessment of their level of clustering relative to random distribution. To eliminate spurious clusters generated by chance, it uses the K-function introduced by \citet{Ripley1976,Ripley1979} to assess the authenticity of the identified clusters. To mitigate contamination from field stars, pyUPMASK employs a Gaussian uniform mixture model. Lastly, it assesses the probability of cluster membership using kernel density estimation (KDE). The reliability and effectiveness of pyUPMASK has been demonstrated by its successful applications in \citet{2022RAA....22e5022B} and \citet{2023ApJS..265...12Q}.


We calculated the member probabilities of the samples by applying the pyUPMASK algorithm on 
{$\it Sample~2$}, {$\it Sample~3,$} and {$\it Sample~4$}. We combined the member star 
candidates selected by \textsc{StarGO} and the results of pyUPMASK using member
probabilities to obtain member star candidates. We identified the open cluster in the NGC\,2546 
region as NGC\,2546 (pink). The Pismis\,5 region contained Pismis\,5 (blue), two neighboring 
clusters, and a separate cluster. We labeled these two clusters Pismis$\,5_{A}$ (red) and Pismis$\,5_{B}$ (green), respectively, and designated another open cluster QZ\,1 (magenta). The clusters in the Alessi\,43 region were Alessi\,43 (orange), Collinder\,197 (purple), and BH\,56 (brown). In all the diagrams below, the blue, red, green, magenta, purple, orange, brown, and pink dots represent Pismis\,5, Pismis$\,5_{A}$, Pismis$\,5_{B}$, QZ\,1, Collinder\,197, Alessi\,43, BH\,56, and NGC\,2546, respectively. 

\subsubsection{Decontamination in CMD}

Performing field-star decontamination in color--magnitude diagrams (CMD) is a crucial step in identifying and characterizing star clusters. To calculate the cluster main sequence lines and their corresponding error positions, we used the Gaussian process-based regression method provided by \citet{2020ApJ...901...49L}. Data points are considered outliers if they deviate from the main sequence band by more than 3$\sigma$, being either bluer than the main sequence line or redder than the equal-mass-binary curve, where $\sigma$ is the uncertainty caused by photometric errors and main sequence line errors. We remove outliers on both sides of the main sequence band and
obtain candidate member stars for eight clusters. The resulting CMDs, as shown in Figure~\ref{fig:age}, exhibit a low-noise main sequence after decontamination. Numbers for the cluster members are displayed in the second column of Table~\ref{tab:table1}.

\subsection{Comparison with previous catalogs} \label{subsec:compare}

\citet{2020A&A...633A..99C} give the member candidates of 2017 open clusters based on {\it Gaia} DR2, all 
of which have member probabilities of greater than or equal to 0.7. 
To maintain the same contamination rate, we also restrict the probability of member stars from 
\citet{2020A&A...633A..99C} to be greater than or equal to 0.9.
We performed a cross-match\footnote{Our cross-match bases on $\alpha$ and $\delta$, and if the difference between the coordinates of the two stars is less than one arcsecond, 
then they are considered as the same star with TOPCAT \citep{tayl05}.} between our clusters and the 
open-cluster catalog given by \citet{2020A&A...633A..99C}. We matched to 53, 0, 0, 0, 168, 
119, 56, and 34 member stars for the clusters Pismis\,5, Pismis$\,5_{A}$, Pismis$\,5_{B}$, QZ\,1, 
Alessi\,43, Collinder\,197, BH\,56, and NGC\, 2546, respectively. In the list of \cite{2020A&A...633A..99C}, 
Pismis\,5, Alessi\,43, Collinder\,197, BH\,56, and NGC\,2546 have 57, 196, 161, 62, and 35 member candidates, 
respectively. These parameters are listed in the third column of the Table ~\ref{tab:table1}.
Comparing our member candidates to the  member candidates given by \cite{2020A&A...633A..99C}, we discover that there is good agreement between the two member-candidate lists. There are also member candidates found by \cite{2020A&A...633A..99C} that are not in our list. We find that the majority of these stars do not meet our criteria because of their inferior photometric quality.

The open clusters 
OC-0467, OC-0469, and OC-0472 discovered by \citet{2022A&A...660A...4H} are in the region of 
Pismis\,5. After matching, Pismis$\,5_{A}$ has 57 member stars that are members of 
OC-0467, 15 member stars of OC-0469, and 1 member star of OC-0472, whereas Pismis$\,5_{B}$ has 7 
member stars that are members of OC-0467, 1 member star of OC-0469, and 2 member stars of OC-0472. 
Even though we employ two clustering methods in our work, we are unable to find OC-0467, 
OC-0469, or OC-0472. We call these two star clusters Pismis$\,5_{A}$ and Pismis$\,5_{B}$. 
We have not found any known clusters sharing members with QZ\,1, and therefore consider it to be a new open cluster.

\begin{table*}[th]
    \centering
    \tiny{
    \caption{{Parameters of eight open clusters in this paper.}}
    \label{tab:table1}
    }
\tabcolsep=0.04cm
\begin{tabular}{*{22}{c}}
    \hline\hline
  
                 &Cluster        &Number  & $\rm {N_{Cantat}}$ & \multicolumn{3}{c}{${\mu}_{\alpha }cos{\delta }$[mas/yr]} & \multicolumn{3}{c}{$\mu_\delta$[mas/yr]} & \multicolumn{3}{c}{RV[km/s]}    & \multicolumn{3}{c}{Distance[pc]}      &\multicolumn{3}{c}{[Fe/H][dex]}     &{Log\,age[dex]}                       &{$\rm {A_v} [mag]$}                    &{Mass[$\rm {M_{\odot}]}$}\\     
                 &               &        &                    & med & 5$\%$ & 95$\%$                                   & med & 5$\%$ & 95$\%$                  & med & 5$\%$ & 95$\%$                      & med & 5$\%$ & 95$\%$                           & med & 5$\%$ & 95$\%$               & & &\\ 
    \hline
                 &Pismis\,5      &$129$   &57(53)              &-5.50 &-5.88 &-4.92                                        &4.20 &3.64 &4.52                          &17.80 &-34.60 &30.23                          &932.70  &909.42 &949.51                &-0.48&-1.22 &0.05                   &$6.70\pm0.05$                         &$1.87\pm0.02$                          &$197.6\pm0.6$\\
                 &Pismis$\,5_{A}$&$151$   &---                 &-5.76 &-6.12 &-5.38                                        &3.64 &3.09 &3.99                          &21.86 &-7.33  &47.54                          &940.31  &911.32 &948.12                &-0.35&-1.48 &0.15                   &$6.65\pm0.05$                         &$1.56\pm0.16$                          &$173.4\pm2.5$\\
                 &Pismis$\,5_{B}$&$82$    &---                 &-5.44 &-5.88 &-4.90                                        &3.29 &2.97 &3.88                          &17.29 &-43.99 &324.36                          &875.74  &861.77 &885.00                &-0.36&-1.96 &0.21                   &$6.85\pm0.05$                         &$2.10\pm0.01$                          &$89.6\pm0.4$\\
                 &QZ\,1          &$243$   &---                 &-8.12 &-8.51 &-7.70                                        &5.71 &5.40 &6.09                          &27.86 &10.74  &38.04                          &754.14  &729.03 &767.57                &-0.30&-0.86 &0.11                   &$7.75\pm0.05$                         &$0.56\pm0.01$                          &$293.5\pm1.4$\\                 
                 &Alessi\,43     &$279$   &196(168)            &-5.60 &-6.10 &-5.09                                        &3.94 &3.45 &4.53                          &17.96 &-20.70 &44.74                          &946.52  &922.34 &964.77                &-0.34&-2.56 &0.12                  &$6.75\pm0.05$                         &$1.62\pm0.08$                          &$394.8\pm0.2$\\
                 &Collinder\,197 &$179$   &161(119)            &-5.77 &-6.21 &-5.25                                        &4.06 &3.34 &4.53                          &25.20 &-25.33 &41.64                          &952,18  &919.23 &970.38                &-0.30&-0.77 &0.00                   &$6.85\pm0.05$                         &$2.40\pm0.04$                          &$241.2\pm0.8$\\
                 &BH\,56         &$155$   &62(56)              &-5.49 &-5.95 &-5.09                                        &5.42 &4.87 &5.75                          &11.59 &-13.92 &34.92                          &903.65  &881.47 &914.33                &-0.43&-1.18 &0.00                   &$6.85\pm0.05$                         &$1.22\pm0.19$                          &$211.7\pm0.5$\\
                 &NGC\,2546      &$354$   &35(34)              &-3.72 &-4.02 &-3.45                                        &3.95 &3.72 &4.24                          &15.64 &0.22   &29.44                          &933.34  &909.04 &951.81                &-0.35&-1.05 &0.01                   &$8.05\pm0.05$                         &$0.84\pm0.03$                          &$515.9\pm8.8$\\
    \hline\hline               
    \end{tabular} 
    \tablefoot{~Col.~1 -- Names of the open clusters; ~Col.~2 -- The number of the cluster members identified by us; 
            ~Col.~3 -- The number of the cluster members identified by \citet{2020A&A...633A..99C}, with numbers in parentheses indicating the number of member stars in common with our study; ~Col.~4--18 -- The proper motion, RV, distance, and [Fe/H]
            of clusters. The med is the median of them and the columns labeled 5$\%$ and 95$\%$ are
            the bounds of the corresponding confidence intervals. ~Col.~19--21 -- The age, 
            extinction, and total stellar mass of clusters.}
    \end{table*}

\section{General properties} \label{sec:properties}

\subsection{Spatial distribution} \label{sec:distribution}
\begin{figure*}
    \centering
    \includegraphics[width=1.0\textwidth]{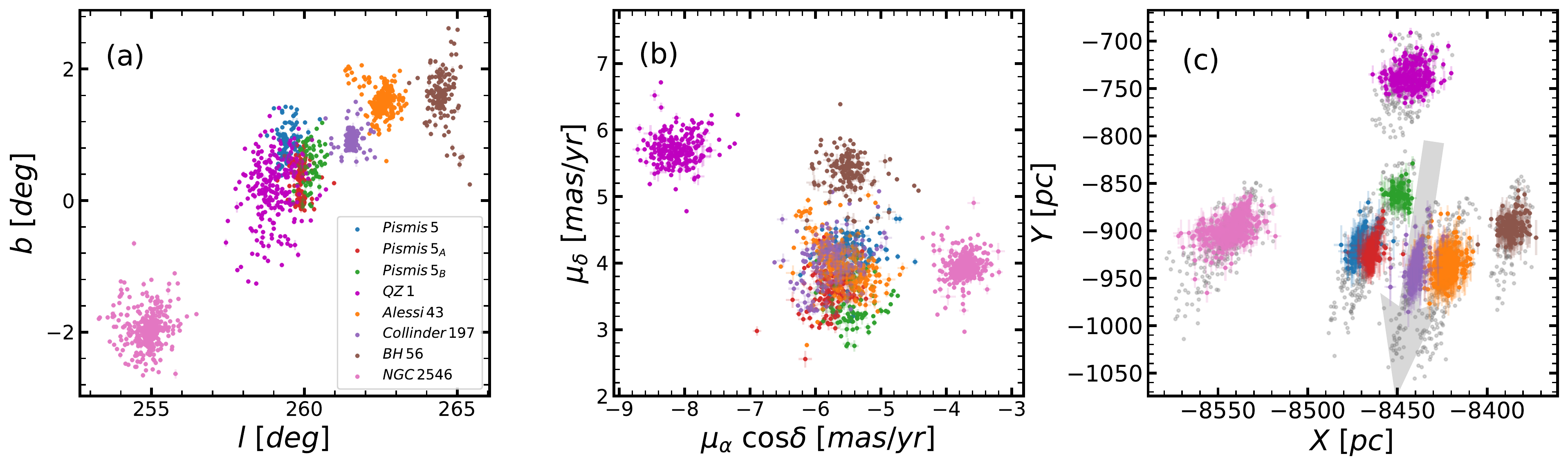}
    \caption{The spatial and proper motion distribution of eight open clusters. ~(a): The spatial distribution of open clusters in the Galactic coordinate system.
    ~(b): The distribution of proper motions of eight open clusters.
    ~(c): The spatial distribution of open clusters in the Galactocentric Cartesian coordinate 
    system on the X-Y plane. The stars with distance correction are shown by color dots, whereas 
    the position of the members without distance correction is shown by grey dots in the background. 
    The direction of the line of sight is shown by the grey arrow. The error bars are depicted in 
    the three diagrams by the grey lines. The data errors in position and proper motion space are 
    much smaller than the scale of the cluster itself. The contribution of errors to cluster search 
    and identification is negligible. \label{fig:distribution}} 
    
    \end{figure*} 

We present the spatial distribution and corresponding errors of eight open clusters in 
Figure~\ref{fig:distribution}~(a). Pismis\,5 ($l=259.36$ deg, $b=0.92$ deg), 
Pismis$\,5_{A}$ ($l=259.90$ deg, $b=0.41$ deg), Pismis$\,5_{B}$ ($l=260.19$ deg, $b=0.54$ deg),
QZ\,1 ($l=258.96$ deg, $b=0.23$ deg), Alessi\,43 ($l=262.52$ deg, $b=1.49$ deg), 
Collinder\,197 ($l=261.53$ deg, $b=0.94$ deg), and BH\,56 ($l=264.48$ deg, $b=1.60$ deg) are located
in the VMR. NGC\,2546 ($l=254.91$ deg, $b=-1.98$ deg) is situated outside the VMR.
The member candidates of Pismis\,5, Pismis$\,5_{A}$, Pismis$\,5_{B}$, and 
QZ\,1 are indistinguishable in the $l-b$ plane. Alessi\,43 and Collinder\,197 also have member candidates that are indistinguishable. Except for QZ\,1, BH\,56, and NGC\,2546, other star clusters 
are mixed and difficult to distinguish in proper-motion space, as seen in 
Figure~\ref{fig:distribution}~(b). The information above suggests that the five clusters are 
possibly a large cluster aggregate, in which Pismis\,5, Pismis$\,5_{A}$, and Pismis$\,5_{B}$ 
form one set of sub-aggregate candidates, and Alessi\,43 and Collinder\,197 form the other.

We studied the distribution of clusters in 3D space, and Figure~\ref{fig:distribution}~(c) gives 
the distribution of clusters in the X-Y plane. Every open cluster shows an apparent 
elongation along the line of sight. The extension is an artificial one caused by directly inverting the
parallaxes to obtain the distances. Because the parallax error has a symmetric 
distribution function, introducing a skewed distribution of errors on the distance obtained by taking the reciprocal leads to a bias in estimating the distance \citep{2019A&A...627A.119C,2020ApJ...889...99Z,2021ApJ...912..162P,2022ApJ...931..156P}. 
To alleviate this artificial elongation, previous studies (e.g., \citealt{2021ApJ...912..162P, 2021AJ....162..171Y, 2022ApJ...931..156P}) have employed the Bayesian approach outlined in \cite{2015PASP..127..994B} and \citep{2019A&A...627A.119C}. And we also follow their method and adopt the prior consist of the density of the field stars and the cluster members in \citet{2021AJ....162..171Y}. 

\begin{equation}
    P\left( d \right) \propto C \cdot d^{2} e^{-\frac{d}{8 \mathrm{[kpc]}}} + \left( 1-C \right) \cdot \frac{1}{\sqrt{2\pi \sigma_{d}^{2}}}e^{-\frac{(d-d_{0})^2}{2 \sigma_{d}^{2}}},
  \end{equation}
where $d_{0}$ is the predicted mean distance of cluster members, adopted as the inverted median parallaxes of clusters. $\sigma_{d}$ is the deviation of the distances between member stars and their cluster center. The coefficient $C$ is the contamination fraction. 
Under the assumption of the Gaussian parallax uncertainties, the likelihood is 
\begin{equation}      
    P( {\varpi}|d) \propto  e^{ - ({\varpi}  - \frac{1} {d} ) ^{2}/2  \sigma_{\varpi} ^{2} },  
  \end{equation}
where $\sigma_{\varpi} ^{2}$ equals the quoted error bar on each parallax measurement, as used in \citet{2019A&A...627A.119C}.
In Figure~\ref{fig:distribution}~(c), the member stars with distance correction are more compactly 
distributed along the line of sight direction. Here we note that the distances through Bayesian parallax inversion 
effectively mitigate the elongation of open clusters. The median, 5$\%$, and 95$\%$ 
quantiles of corrected distances for open clusters are listed in Table~\ref{tab:table1}.

We get more precise 3D positions for the cluster member candidates after correcting for distance. Furthermore, we carry out Monte Carlo (MC) simulations to estimate the uncertainty in the 3D position
transformed by $\alpha$, $\delta$, and distance. With the initial parameters serving as the mean and 
the associated uncertainties serving as the standard deviation, we resampled the $\alpha$, $\delta$, and 
distance of each source from a Gaussian distribution. We used 1000 resampled data to perform 
the coordinate conversion for each source, yielding 1000 3D positions for each star. The 
uncertainties of each star position were determined using the standard deviation of 1000 results. 
Due to the substantial distance uncertainties brought on by the parallax uncertainties, the 
uncertainties in the 3D position are greater than for the 2D ones.

In proper motion space, five open clusters are indistinguishable, but 
in the X-Y plane, we can see that the eight open clusters are distinct from one another. As a result, 
it can be inferred that Pismis\,5, Pismis$\,5_{A}$, and Pismis$\,5_{B}$ are three separate open clusters, 
and the overall region is a triple open-cluster candidate. Alessi\,43 and Collinder\,197 is also a binary open-cluster candidate.

\subsection{Age determination} \label{subsec:age}

\begin{figure*}
    \includegraphics[width=0.8\textwidth]{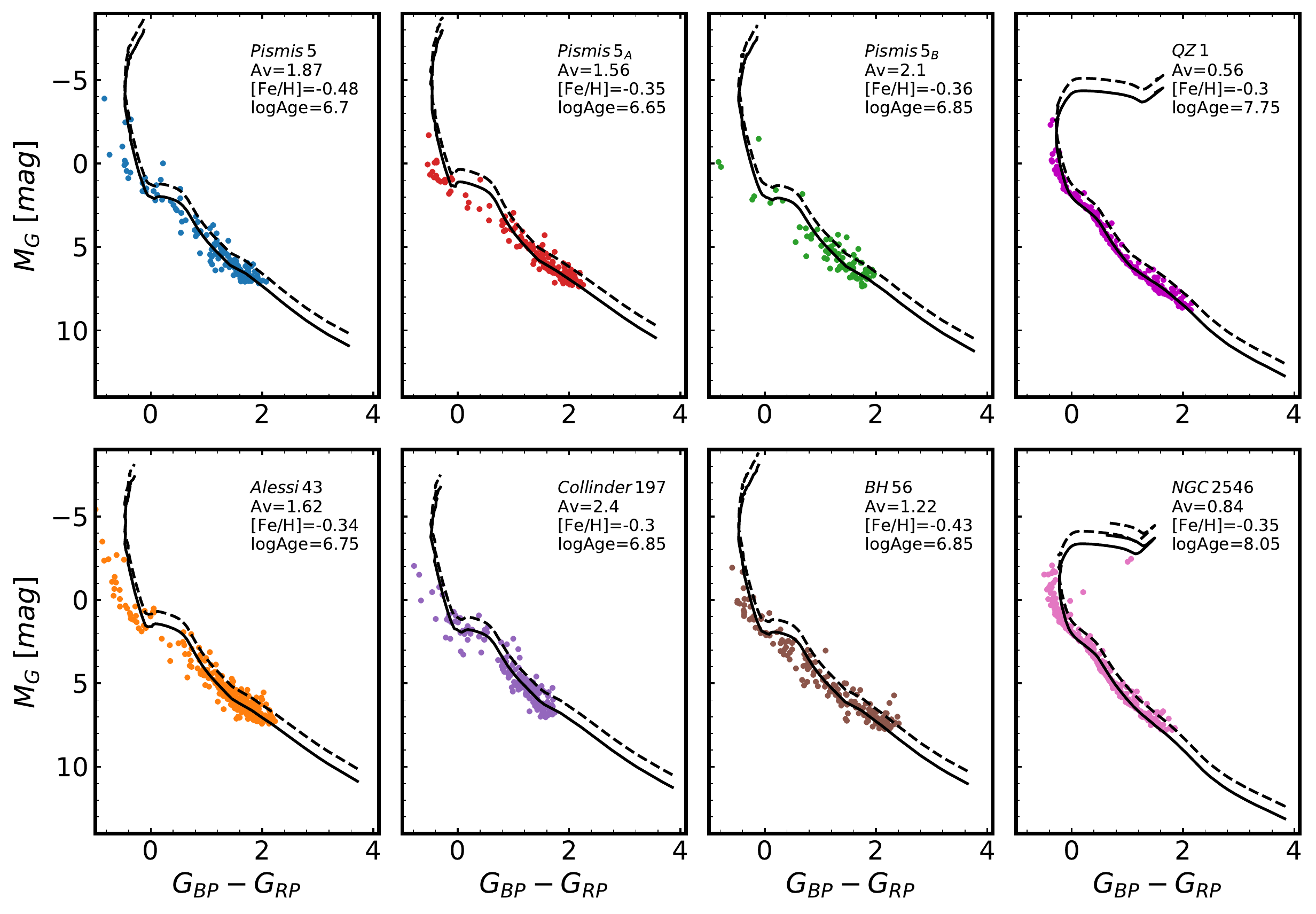}
    \centering
    \caption{CMDs of Pismis\,5, 
    Pismis$\,5_{A}$, Pismis$\,5_{B}$, QZ\,1, Alessi\,43, Collinder\,197, BH\,56, and NGC\,2546.
    The solid lines are the isochrones from PARSEC and the corresponding parameters are marked in 
    the figure. The dashed lines are the curve of equal-mass binaries.
    \label{fig:age}}
    \end{figure*}   
The CMD is also a common tool used to estimate the basic parameters of open 
clusters, including reddening, metallicity, and age. We present the CMDs of eight open clusters in 
Figure~\ref{fig:age}. Only QZ\,1 and NGC\,2546 exhibit clear main sequence tracks; the other 
clusters have minor main sequence dispersion due to their location in molecular clouds, which are 
more extinct and younger.

{\it Gaia} DR3 provides [Fe/H] values based on synthetic spectrum templates for main 
sequence stars of spectral types F-G-K. Therefore, we only use stars of spectral types 
F-G-K (effective temperatures between 3500 and 7500 K) as metallicity samples. We consider the medians 
of these samples to be the metallicity of the cluster for the isochrone fit. Table~\ref{tab:table1} 
provides the median, 5$\%,$ and 95$\%$ quantiles of metallicities. We use the PARSEC stellar evolution isochrones \citep{2012MNRAS.427..127B,2014MNRAS.444.2525C} for the Gaia filters, to fit the isochrones of open clusters with an age range of 1 Myr to 1 Gyr, at steps of log age=0.05.
We use the following function \citep{2019ApJS..245...32L} to determine the best-fitting isochrone and obtain the value of $A_V$:

\begin{equation} 
    \centering
    \overline{d}\ ^{2} = \sum_{k=1}^{n} | \ \textbf{{x}}_{k} - \textbf{{x}}_{k, nn} \ |^{2} / n,
    \end{equation}
where $n$ is the number of members, and 
$\textbf{{x}}_{k} = [{\it G}_{k} + \Delta_{\rm G} + {\it A}_{\rm G}, 
\ ({\it G}_{\rm BP} - {\it G}_{\rm RP})_{\it k} + ({\it A}_{\rm BP} 
- {\it A}_{\rm RP})]$, $\textbf{{x}}_{k, nn}$ is the position of the 
$k$th member star and the nearest neighboring point to this star in the
isochrone, respectively. The nearest neighboring point is searched for using the K-D Tree method.

The relations between $A_{\rm G}$, $A_{\rm {G_{BP}}}$, and $A_{\rm {G_{RP}}}$ are calculated using the  following
formula:

\begin{equation}
  \begin{split}
A_{M} / A_{\rm V} = & c_{1 M}+c_{2 M}\left(G_{\rm B P}-G_{\rm R P}\right)+c_{3 M}\left(G_{\rm B P}-G_{\rm R P}\right)^{2}+\\
                    & c_{4 M}\left(G_{\rm B P}-G_{\rm R P}\right)^{3}+c_{5 M} A_{\rm V}+c_{6 M} A_{\rm V}^{2}+\\
                    & c_{7 M}\left(G_{\rm B P}-G_{\rm R P}\right) A_{\rm V},
  \end{split}
\end{equation}
where the subscript $M$ indicates the $G$, $G_{\rm BP}$, and $G_{\rm RP}$ band, and $c_{1...7M}$ belong to a set of 
coefficients defined in Table\,1 of \citet{2018A&A...616A...1G}. We used these equations to 
get the square of the mean separation between each isochrone and its member stars. 
The Nelder-Mead algorithm, which is part of the "scipy" package \citep{2020NatMe..17..261V}, is  
used to determine the optimal age and $\rm {A_V}$. The isochrones are displayed in Figure~\ref{fig:age} 
and the corresponding parameters are listed in Table~\ref{tab:table1}. The 
uncertainty in the fitted age for each cluster is considered to be twice the age step.

With the exception of NGC\,2546 and QZ\,1, the clusters are young and have similar ages.
According to their ages, NGC\,2546 is thought to have formed first, followed by QZ\,1, Pismis\,5, Collinder\,197, BH\,56, and Pismis$\,5_{B}$, with Pismis$\,5_{A}$ and Alessi\,43 forming last in that sequence.
This phenomenon is called age spread \citep{2003ARA&A..41...57L}, and is due to the time needed for 
star formation and cluster gestation.
The similar ages of Pismis\,5, Pismis$\,5_{A}$, and Pismis$\,5_{B}$ indicate that they are coeval.
Pismis\,5, Pismis$\,5_{A}$, and Pismis$\,5_{B}$ have similar locations, kinematics, and metallicities,
have young ages, and show little difference in age, implying that they are a primordial triple open-cluster candidate.
Similarly, Alessi\,43 and Collinder\,197 are a primordial binary open-cluster candidate.

\subsection{Mass function and mass segregation} \label{sec:mass}

\subsubsection{Mass function} \label{sec:mf}

\begin{figure}
    \centering
    \includegraphics[width=0.50\textwidth]{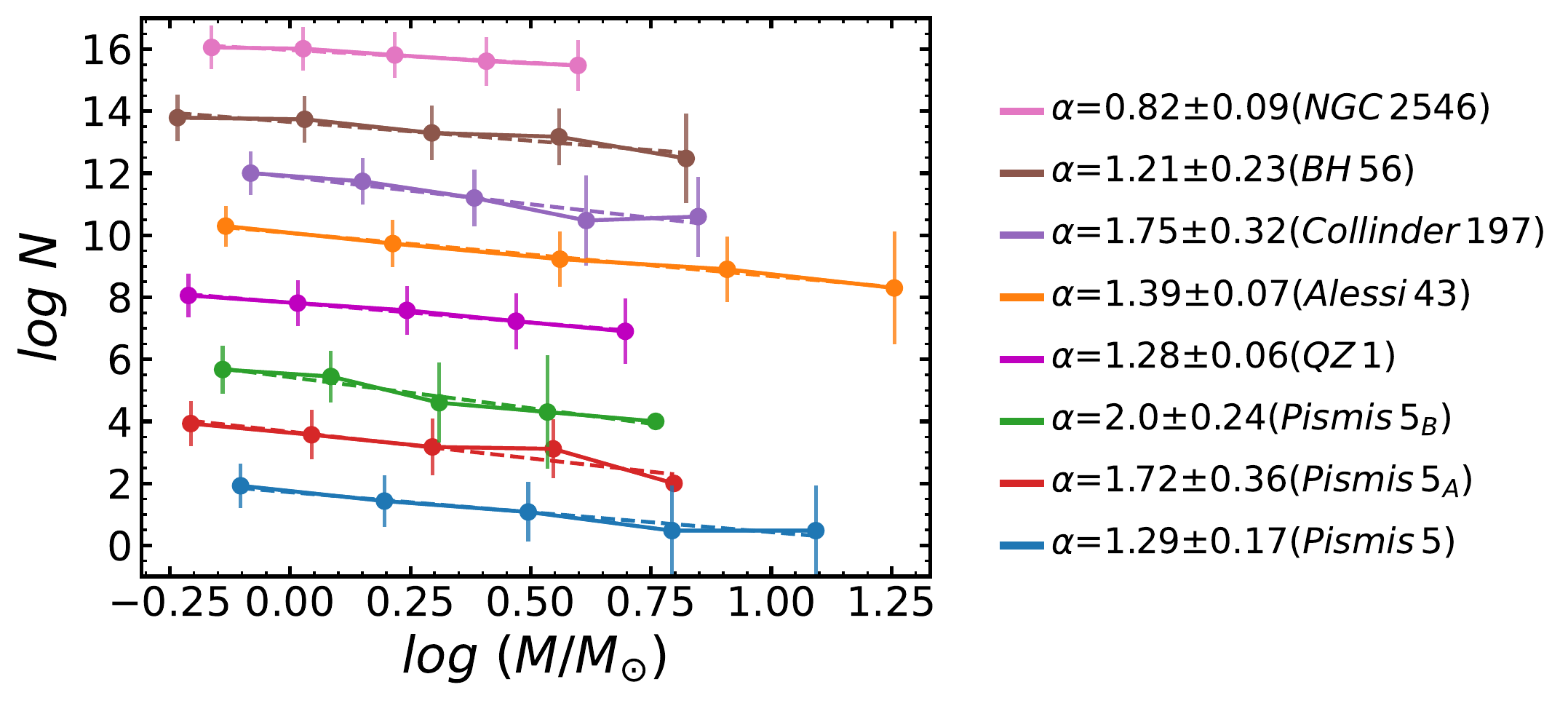}
    \caption{Present-day mass function of Pismis\,5, 
    Pismis$\,5_{A}$, Pismis$\,5_{B}$, QZ\,1, Alessi\,43, Collinder\,197, BH\,56, and NGC\,2546. The stellar mass is assigned to each source by nearest-neighbor interpolation in the Gaia colour–magnitude space ($BP-RP$ versus $M_G$) using the PARSEC isochrones. The Y-axis is shifted manually by 2 units so that the mass function of each cluster is visible. The dashed lines are the fitted mass functions and the quantity $\alpha$ is the fitted slope. The error bars represent Poissonian uncertainties ($\frac{1}{\sqrt{N}}$) \citep{2019MNRAS.482.1471B}.}
    \label{fig:mf}
    \end{figure}

One of the most fundamental properties of an open cluster is its stellar mass. The stellar mass of 
members is estimated via the mass--magnitude relation of the PARSEC stellar-evolution isochrone,
and we obtain stellar mass for each member candidate of eight open clusters. Table~\ref{tab:table1} also 
shows the total stellar mass of Pismis\,5, Pismis$\,5_{A}$, Pismis$\,5_{B}$, QZ\,1, Alessi\,43, 
Collinder\,197, BH\,56, and NGC\,2546. The mass function (MF) is the most basic distribution
of clusters \citep{2019ARA&A..57..227K}. The functional form of the mass 
function is:
\begin{equation} 
    \centering
    dN/dm\propto m^{-{\alpha}},
    \end{equation}
where $N$ is the number of member stars within the mass bin $dm$, and $\alpha$ is the slope of the mass function \citep{1955ApJ...121..161S}.
The members are incomplete beyond $\sim$ 18 mag. In order to obtain a more accurate mass function, 
we select stars with apparent magnitudes smaller than or equal to 18 mag. 
By using the linear least-squares fitting, we are able to determine the values of $\alpha$ for 
the mass functions shown in Figure~\ref{fig:mf}.

The mass functions of the eight clusters exhibit varying degrees of smoothness, with NGC\,2546 having the smoothest mass function, followed by BH\,56, QZ\,1, Pismis\,5, Alessi\,43, and Pismis$\,5_{A}$, while Collinder\,197 and Pismis$\,5_B$ have steeper mass functions.
The proportion of low-mass stars decreases with a flatter mass function. 
On the one hand, we note that these clusters are far from the Sun and the low-mass stars may be too 
faint to be observed. As a result, the fraction of low-mass stars from clusters is affected by data 
incompleteness for the cluster. On the other hand, both internal and external forces have an impact on
the loss of less massive stars. The less massive stars migrate to the outer regions of the cluster as they evolve due to internal relaxation, making them more vulnerable to being stripped away by Galactic tidal field.

The cluster NGC\,2546, which is not part of the VMR, has the flattest mass function as many of its 
low-mass stars have already been lost to the Galactic tidal field.
Similarly, BH\,56, QZ\,1, Pismis\,5, and Alessi\,43 are located at  
the periphery of the VMR region and have a flatter mass function in comparison to the clusters in the 
interior of the VMR (Pismis$\,5_{A}$, Pismis$\,5_{B}$, and Collinder\,197), as they are more vulnerable to the loss of low-mass stars by external perturbations.

\subsubsection{Mass segregation} \label{sec:ms}
\begin{figure*}
    \centering
    \includegraphics[width=0.8\textwidth]{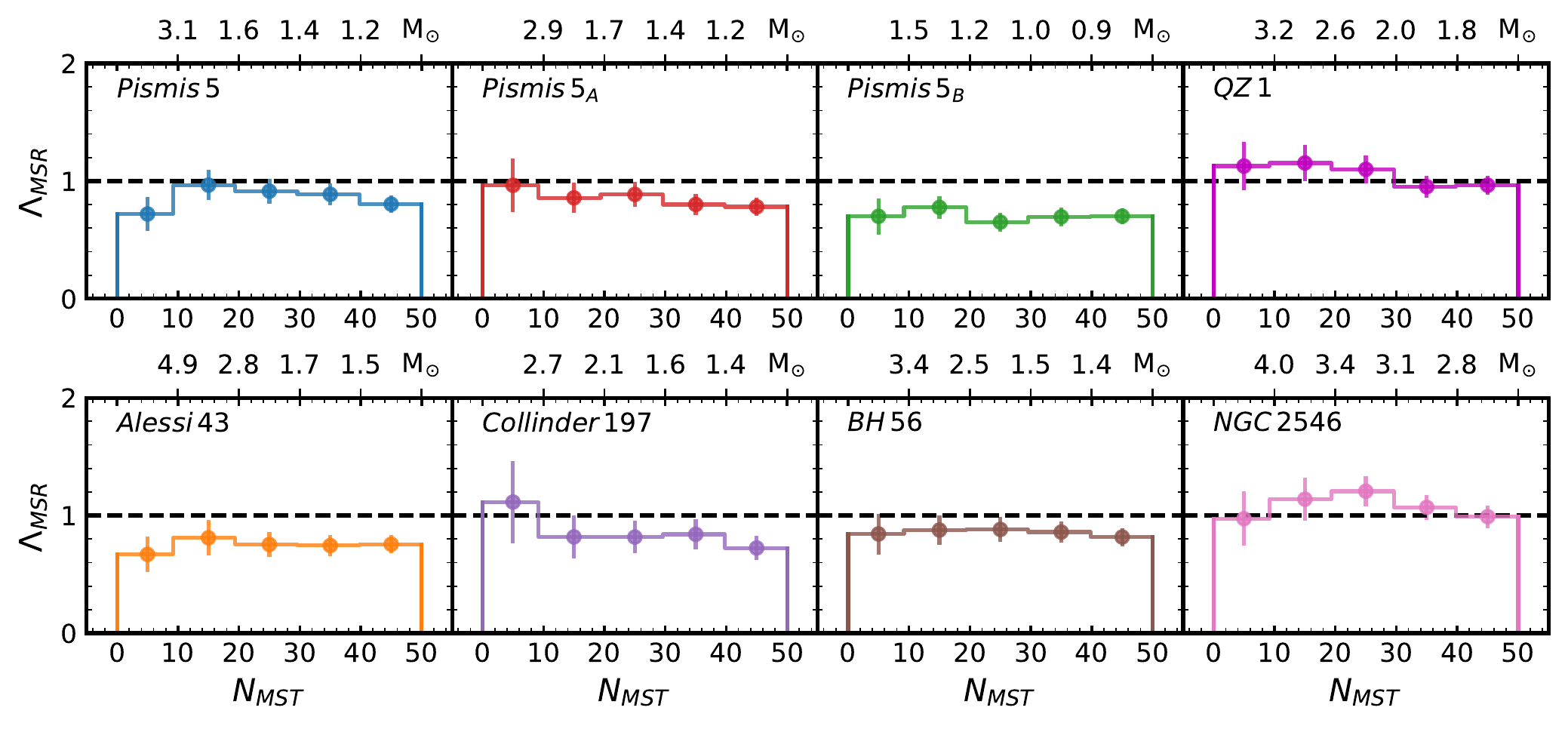}
    \caption{ ``Mass segregation ratio'' ($\Lambda_{\rm MST}$) for the 50 most massive 
    members with a bin size of 10 stars in the 3D spatial position of Pismis\,5, 
    Pismis$\,5_{A}$, Pismis$\,5_{B}$, QZ\,1, Alessi\,43, Collinder\,197, BH\,56, and NGC\,2546.
    The dashed line ($\Lambda_{\rm MST} = 1$) indicates an absence of mass 
    segregation. The increasing value of $\Lambda_{\rm MST}$ indicates a more significant 
    degree of mass segregation. The error bars indicate the uncertainties obtained 
    from 100 realizations of ${l}_{\rm normal}$.
    \label{fig:ms}}
    \end{figure*}

Mass segregation is frequently observed in open clusters.
We use the minimum spanning tree method provided by \cite{2009MNRAS.395.1449A}
to measure the degree of mass segregation of eight open clusters. This method determines the length of 
the minimum spanning tree of the N most massive stars as $l_{\rm massive}$, the average length of 
the minimum spanning tree of sets of N random stars as $l_{\rm norm}$, and the standard deviation of 
the lengths $l_{\rm norm}$ as $\sigma_{\rm norm}$. \cite{2009MNRAS.395.1449A} define the ``mass 
segregation ratio (MSR)'' ($\Lambda_{\rm MSR}$) as the ratio between the average random 
path length and that of the massive stars:

 \begin{equation}
{\Lambda}_{\rm MSR} ={\frac{  \langle  l_{\rm norm}  \rangle }{ l_{\rm massive}}} \pm {\frac{  \sigma _{\rm norm}}{ l_{\rm massive}}},
    \end{equation}
where $\Lambda _{\rm MSR}$ significantly $ > 1$ indicated mass segregation, and $\Lambda _{\rm MSR}$ 
$\leq  1$ showed no mass segregation. To quantify the mass segregation of the 
50 most massive members of the clusters, we divide these stars into five groups and use 
the Python package Minimum Spanning Trees (Mistree) \citep{2019JOSS....4.1721N} to calculate the mass segregation 
ratios of these five groups. The minimum spanning tree method is applied using the 3D 
positions of the members, and the results are displayed in Figure~\ref{fig:ms}.
Only ten stars of Collinder\,197 show mass segregation, but with large uncertainty, suggesting that mass segregation is not present in this cluster. Our analysis reveals that Pismis\,5, Pismis\,$5_A$, Pismis\,$5_B$, Alessi\,43, and BH\,56 do not exhibit mass segregation. Therefore, among the eight target open clusters, only NGC\,2546 and QZ\,1 show significant evidence of mass segregation.

Mass segregation in star clusters may be either primordial or dynamical. The 
hierarchical formation scenario postulates that open clusters form in small clumps, which merge 
to form larger mass-segregated systems in young clusters. Dynamical mass segregation is 
the consequence of the equipartition of energy during the two-body encounters. N-body 
simulations suggest that mass segregation occurs on a timescale of a few million years \citep{2022A&A...659A..59T}.
QZ\,1 and NGC\,2546 have ages of 56.23 and 112.20 Myr. It is very likely that they have lost the 
imprint of substructures from their parent molecular clouds. Therefore, the mass segregation 
of QZ\,1 and NGC\,2546 is probably caused by the equipartition of kinetic energy via two-body relaxation,
which makes massive stars within clusters move toward their center, whereas lighter stars move toward the 
outskirts. The other clusters do not exhibit mass segregation, which is probably because they are younger and do not have enough time to dynamically evolve into a state of mass segregation.


\subsubsection{Initial stellar mass} \label{sec:imf}
\begin{figure*}
    \centering
    \includegraphics[width=1.0\textwidth]{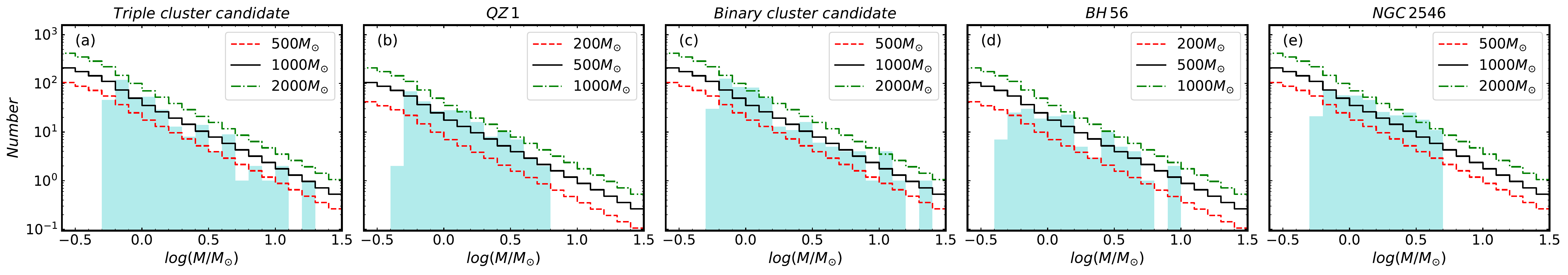}
    \caption{~ Initial stellar mass estimation of the aggregate candidate Pismis\,5, 
    Pismis$\,5_{A}$, and  Pismis$\,5_{B}$, (a); QZ\,1 (b), the aggregate candidate composed of Alessi\,43 and 
    Collinder\,197; (c) BH\,56 (d); and NGC\,2546 (e) based on the IMF model \citep{2001MNRAS.322..231K}. 
    The histograms, which are fitted with IMFs of varied total mass, show the observed mass 
    distribution.}
    \label{fig:imf}
    \end{figure*}

The evolution of a cluster is the result of cluster internal dynamics, gas expulsion, stellar 
evolution from a realistic initial stellar mass function, and the tidal field of the host 
galaxy \citep{2022A&A...660A..61D}. Therefore, the initial stellar mass is an essential parameter
for the cluster. We removed data beyond 18 mag to estimate the initial stellar mass of the cluster, because the data fainter than 18 mag are incomplete. As 
Pismis\,5, Pismis$\,5_{A}$ and Pismis$\,5_{B}$ form a triple open-cluster candidate, we evaluate their initial 
stellar mass as a whole. Likewise, Alessi\,43 and Collinder\,197 are also considered as a whole. 
We use the initial mass function (IMF) given by \citet{2001MNRAS.322..231K}. The multiple-part power-law IMF is represented by the following formula,

\begin{equation}
    \xi(m) \propto m^{-\alpha_{i}}, 
    \quad \left\{
    \begin{aligned}
    \alpha_{1} &=+1.8 \pm 0.5, \quad 0.08 \leq m / \mathrm{M}_{\odot}<0.50,\\
    \alpha_{2} &=+2.7 \pm 0.3, \quad 0.50 \leq m / \mathrm{M}_{\odot}< 1.00\\
    \alpha_{3} &=+2.3 \pm 0.7, \quad 1.00 \leq m / \mathrm{M}_{\odot}\le50.00, \\
    \end{aligned}
    \right.
    \end{equation}
where $\xi(m)\,dm$ is the number of stars in the mass interval $m$ to $m+dm$, and ${\alpha}$ is the slope 
of the mass functions, the value of which varies with stellar mass.
The lower and upper limits of the IMF are 0.08 $\rm M_{\odot}$ and 50 $\rm M_{\odot}$, respectively.
We constructed mass-distribution histograms for the eight open clusters and used the IMF to create a series of mass histogram profiles with different initial masses, as shown in Figure~\ref{fig:imf}.
BH\,56 and QZ\,1 have an initial stellar mass of about 500 $\rm M_{\odot}$, while the triple open-cluster candidate, the binary open-cluster candidate, and NGC\,2546 all have an initial stellar mass of approximately 1000 $\rm M_{\odot}$.

We estimate that the triple and binary open-cluster candidates have lost roughly 54$\%$ and 36$\%$ of 
their stellar mass in comparison to the present-day stellar mass of clusters.
The two older clusters, NGC\,2546 and QZ\,1, show significant mass segregation,
losing mostly low-mass stars in the peripheral at moderate mass-loss rates of 48$\%$ and 
41$\%$, respectively. BH\,56 has a stellar mass-loss rate of 56$\%$, because it is situated at the 
boundary of the VMR in a complicated external environment that is prone to star loss. As these 
clusters reside in molecular clouds with significant extinction, the completeness of their member stars is greatly affected. The incompleteness of the member stars can lead to an underestimation of the initial mass. Therefore, the actual mass-loss rate ought to be higher than the one calculated now. 
Open clusters, as they evolve and keep losing stars, contribute to the field stellar populations; full disintegration happens when they can no longer be distinguished from the field in terms of stellar density.

\section{Discussion} \label{sec:discussion}

\subsection{Kinematic analysis} \label{sec:kinematics}
\begin{figure*}[ht]
    \centering
    \includegraphics[width=1.0\textwidth]{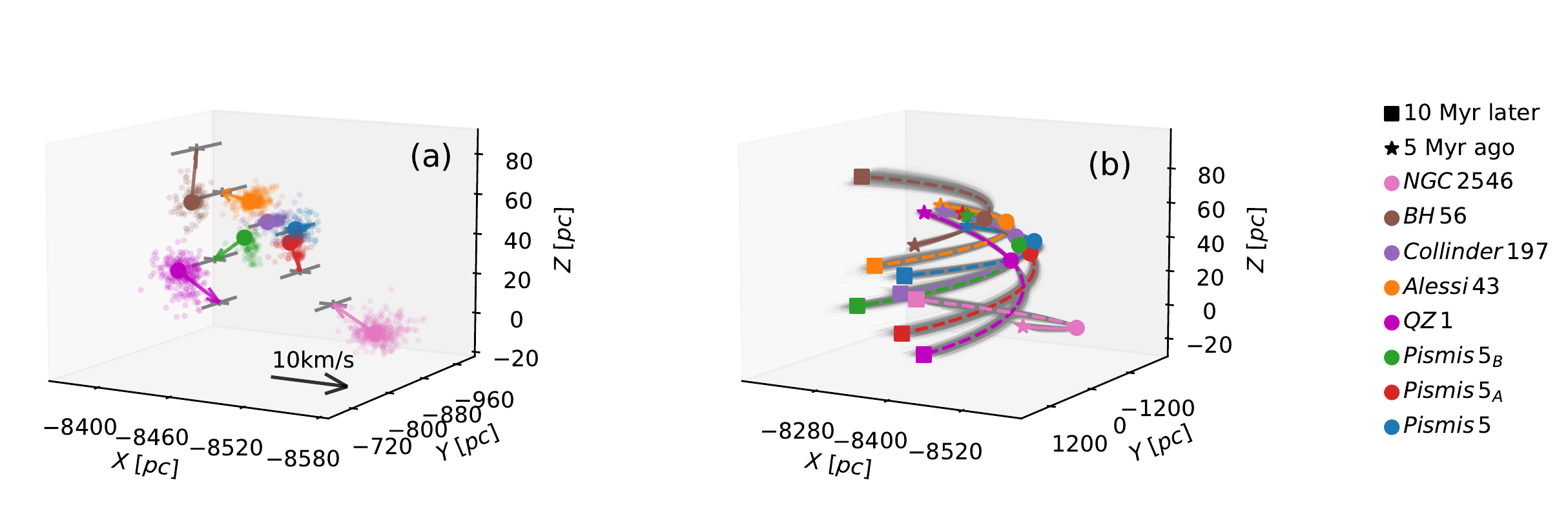}
    \caption{Relative motions and orbits of eight open clusters. ~(a) The colored dots represent the members of eight open clusters and the larger dots 
    represent the centers of the clusters. The directions of the arrows for Pismis$\,5_{A}$, 
    Pismis$\,5_{B}$, QZ\,1, and NGC\,2546 indicate the direction of motion with respect to Pismis\,5. The 
    length of the arrows represents the magnitude of the velocity. The arrow directions for
    Alessi\,43 and BH\,56 represent the direction of movement relative to Collinder\,197, and the length 
    of the arrow represents the magnitude of the speed. The gray error bars represent the velocity 
    uncertainties in the X, Y, and Z directions. ~(b) The orbits of eight open clusters. 
    The stars display the position before 5 Myr, whereas the solid lines depict the orbits from 
    before that time to the present. The cluster positions after 10 Myr are indicated by the squares, 
    while the dotted lines depict the trajectory from the present to that point. The 
    gray trajectory is the result of 1000 simulations, meaning the uncertainties of the orbit.
    \label{fig:kinematics}}
    \end{figure*}



We calculate the 3D velocities (U, V, W) in the Galactocentric Cartesian coordinates with 
Astropy using the proper motions and RVs from {\it Gaia} DR3. We take the median values of 
3D velocities to represent the overall motion of the clusters. We utilize MC simulations (the same steps as procedures to estimate the uncertainties of the 
locations in Section~\ref{sec:distribution}) to estimate the uncertainties in the 3D velocities, and these are 
shown as gray error bars in Figure~\ref{fig:kinematics}~(a).
The velocity errors are relatively small in the X and Z directions 
and large in the Y direction. But within the error range, the relative motion trend of the cluster 
remains unchanged.

We split the eight clusters into two groups 
in order to analyze the relative motion of the clusters. The first group includes Pismis\,5, 
Pismis$\,5_{A}$, Pismis$\,5_{B}$, QZ\,1, and NGC\,2546, because QZ\,1 and NGC\,2546 are closer 
to the triple open cluster candidate. We obtain the relative motion of these four clusters with respect to Pismis\,5.
As shown in Figure~\ref{fig:kinematics}~(a), Pismis$\,5_{A}$, Pismis$\,5_{B}$, and 
QZ\,1 will be far away from Pismis\,5, indicating that members in the triple open-cluster candidate may 
gradually disperse. While NGC\,2546 moves towards Pismis\,5, it may gradually approach the triple open-cluster candidate in the future. 
The second group is made up of Alessi\,43, Collinder\,197, and BH\,56. The motions of Alessi\,43 and
BH\,56 relative to Collinder\,197 are shown in Figure~\ref{fig:kinematics}~(a). Alessi\,43 and BH\,56 
move away from Collinder\,197, with BH\,56 moving away from Collinder\,197 more rapidly due 
to its higher velocity. The two aggregate candidates are also found to be separated from each other. The seven clusters in the VMR have a tendency to separate, and the NGC\,2546 shows a trend towards moving closer to the VMR.


We use the Python \texttt{galpy} package \citep{2015ApJS..216...29B} \footnote{The Galactic 
gravitational potential used here is "\texttt{MWPotential2014}", a model that comprises the bulge, 
disk, and halo. Parameters of the model were fitted to published dynamical data of the Milky Way.} to 
calculate orbits of star clusters based on the data of position, proper motion, and RVs. As 
most of our open clusters are young, we backtrack their orbital motion by 5 Myr. In order to ensure the 
accuracy of the orbits, we only calculate the orbit of the next 10 Myr. To calculate the orbital 
uncertainty, we employ the MC simulation described in Section~\ref{sec:distribution}. As shown in Figure~\ref{fig:kinematics}~(b), the gray orbits represent the uncertainties 
in the cluster orbit.

After 10 Myr, BH\,56 will be the cluster that is the furthest away from the other seven, possibly due to its 
high relative velocity. The NGC\,2546 will be getting closer and closer to the binary and triple open-cluster candidates.
Pismis\,5, Pismis$\,5_{A}$, Pismis$\,5_{B}$, QZ\,1, Alessi\,43, and Collinder\,197 have similar orbits.
The separation between these six clusters is projected to gradually increase over the next 10 Myr. 
The members of the binary and triple open-cluster candidates are expected to progressively separate due to their orbital motions.
This is in accordance with the star cluster kinematics tendency to move apart from one another.

\subsection{Double/three or binary/triple?} \label{sec:binary}

Star clusters with slight angular separations are identified as double clusters. However, not all 
double clusters are binary clusters because a binary cluster means that the objects are gravitationally 
bound in a physical pair. The double clusters resulting from resonant trapping or hyperbolic 
encounters may share common kinematics or occupy a limited space but are not physical 
binary clusters. 


There are several mechanisms for the formation of binary and multiple clusters. The cluster aggregates may have formed simultaneously in the same molecular cloud or stellar complex. Alternatively, the collapse of a nearby cloud induced by stellar winds or supernova shocks generated by one open cluster may trigger the formation of companion clusters, ultimately forming a cluster aggregate \citep{2021A&A...649A..54P}. These two processes create cluster aggregates, which are referred to as primordial cluster aggregates. Primordial cluster aggregates share similar spatial positions, kinematics, ages, and metallicities. Moreover, the cluster aggregates may also be formed completely separately and then be captured together by tidal forces. Generally, the captured cluster aggregates have similar spatial positions and kinematics but different ages and metallicities \citep{2009A&A...500L..13D}.

We calculate the median of the RVs of each cluster to represent the overall RV of the cluster. 
And we compute 5$\%$ and 95$\%$ quantiles to provide the 90$\%$ central confidence interval for RVs in Table~\ref{tab:table1}. 
The RVs for Pismis\,5, Pismis$\,5_{A}$, and Pismis$\,5_{B}$ are
approximately 17.80 km/s, 21.86 km/s, and 17.29 km/s, respectively, exhibiting slight variations among them. Combined with the proper motions, the 
three clusters have similar but not identical kinematics. In addition, these clusters exhibit a similar 3D spatial distribution, young ages, and a minimal difference in metallicity, and therefore they are a primordial triple open-cluster candidate.
Similarly, the RVs of Alessi\,43 and Collinder\,197 are 17.96 and 25.20 km/s, respectively. 
These two clusters are also a primordial binary open-cluster candidate.

According to the findings of the N-body simulation by \citet{2010ApJ...719..104D}, 
the evolution of the original binary cluster can come in one of three forms: merger, shredded secondaries, and separated twins.
\cite{2021MNRAS.506.4603D} also prove that the proportion of binary and multiple clusters is higher in the Galactic environment at 20 Myr in N-body simulation. At 50 Myr, the proportion of merged clusters is higher. This demonstrates that the existence of binary and multiple star clusters is only temporary. Based on the kinematic analysis, we presume that the members of the binary and triple clusters will gradually separate. 
However, the gradual separation 
of multiple cluster members over the next 10 Myr does not mean they will become separate clusters.
\cite{2016MNRAS.457.1339P} presented the orbits of the merging process of star clusters. Clusters can 
be separated from each other first then merge, or revolve and merge. Therefore, the cluster aggregate candidates in this work may merge, 
become separate clusters, or disintegrate.

\subsection{Comparison with other works on VMR} \label{sec:cluster}

The VMR is composed of three adjacent emission peaks, and it has been studied many times.
Investigations of the VMR have focused on its young stellar objects \citep{1994A&AS..104..233P}, the dense cores of Vela-D and Vela-C \citep{2010ApJ...723.1065O,2019A&A...628A.110M}, the {\ion{H}{ii}} regions \citep{2018A&A...617A..63P}, and 3D structure
\citep{2021A&A...655A..68H}, most in association with an investigation of star formation 
in molecular clouds. The two investigations that are most relevant to our work are listed below.

\citet{2009MNRAS.397.1915B} use data from the near-infrared Two-Micron All Sky Survey
(2MASS)\footnote{The 2MASS, All Sky data release \citep{2006AJ....131.1163S} –http://www.ipac.caltech.edu/2mass/releases/allsky/.} 
to identify member stars and investigate its age, radial density profile, mass function, and other properties. These authors estimated an age of $5\pm 4$ Myr for the Pismis\,5 main sequence stars, which is consistent with our result of $5\pm 0.6$ Myr. They also obtained a mass function slope of $-1.03\pm 0.19$ for Pismis\,5, while our result is $-1.29\pm 0.17$.
There are differences between the parameters provided by \citet{2009MNRAS.397.1915B} and those used here, which are most likely due to 
different initial data and different member star determination methods. 
While these latter authors use field decontamination, we use a clustering algorithm that combines 
\textsc{StarGO} and pyUPMASK. The triple cluster we find was discovered using 
5D data, whereas Pismis$\,5_{A}$ and Pismis$\,5_{B}$ are not significantly clustered in 2D position 
space. Field decontamination might miss Pismis$\,5_{A}$ and Pismis$\,5_{B}$ in the $RA-Dec$ plane. It is also 
possible to consider this triple open-cluster candidate as one cluster without considering other dimensions.

By cross-matching near-infrared 2MASS data with the {\it Gaia} DR2 catalog, 
\citet{2021A&A...655A..68H} use the FEDReD (Field Extinction-Distance 
Relation Deconvolver) algorithm to unravel the 3D structure of Vela. These authors obtain a 3D cube of extinction density, as well as measurements of 14 dense extinction clouds and nine cavities. 
We compare the eight clusters to the structures they obtained and find that, 
except for NGC\,2546, the remaining seven clusters are  
in density clump\,1 and clump\,3 obtained by \citet{2021A&A...655A..68H}. The center of clump\,1 is $l=261.74, b=-0.75$ deg, and the distance of it ranges from 750 to 3300 pc. The clump\,1 contains Pismis\,5, Pismis$\,5_{A}$, Pismis$\,5_{B}$, and QZ\,1. The clump\,3 is adjacent to clump\,1 in $l-b$ space and the center of clump\,3 is $l=261.74, b=-0.75$ deg. The distance from clump\,3 to the Sun ranges from 470 to 1660 pc. The clump\,3 contains Alessi\,43, Collinder\,197, and BH\,56.
The positions of our target clusters are consistent with the high-density clouds in their Fig.\,4, which demonstrates the reliability of the clusters we identified.

\section{Conclusions}\label{sec:sum}

Using highly precise astrometric data of {\it Gaia} DR3, we obtained eight clusters using \textsc{StarGO} and pyUPMASK with a membership 
probability of at least 90$\%$. All open clusters are located
in the VMR except for NGC\,2546. One of the eight clusters is
discovered by us for the first time, and we name it QZ\,1.
Pismis\,5, Pismis$\,5_{A}$, and Pismis$\,5_{B}$ are identified as a triple open-cluster candidate because of their 
close distribution in 3D space and their common kinematics. Alessi\,43 and 
Collinder\,197 are also recognized as a binary open-cluster candidate. With the exception of NGC\,2546 and QZ\,1,
all clusters are relatively young and of similar age. 
Therefore, both triple and binary open-cluster candidates are primordial aggregate candidates.
The NGC\,2546 and QZ\,1 have evolved mass segregation, which may be caused by dynamical evolution.
The smoothest mass function of NGC\,2546 may be caused by the loss of low-mass stars due to the 
external environment and the incompleteness of the data.

Through the relative motions of the clusters, we find that Pismis\,$5_{A}$, Pismis\,$5_{B}$, and QZ\,1 have 
a tendency to move away from Pismis\,5, whereas NGC\,2546 is migrating toward the triple open-cluster candidate.
Alessi\,43 and BH\,56 are moving away from Collinder\,197, with BH\,56 exhibiting a higher velocity. These motion trends suggest that the open clusters in the VMR are slowly separating from each other. Moreover, the orbital motions of the clusters confirm that the distance between them is gradually increasing. It is likely that the triple and binary open-cluster candidates may gradually separate or disintegrate in the future.
\begin{acknowledgements} 
The authors thank the reviewer for the very helpful comments and suggestions.
The authors acknowledge the Chinese Academy of Sciences (CAS) "Light of West China" Program, No. 2022-XBQNXZ-013, the Natural Science Foundation of Xinjiang Uygur Autonomous Region, No.2022D01E86,  National Natural Science Foundation of China under grant U2031204, the National Key R$\&$D program of China for Intergovernmental Scientific and Technological Innovation Cooperation Project under No. 2022YFE0126200
and the science research grants from the China Manned Space Project with NO. CMS-CSST-2021-A08.
M.Q. thanks Ali Esamdin, Hubiao Nui, Xianhao Ye, Xiaokun Hou, and Guimei Liu for helpful discussions.
This work has made use of data from the European Space Agency (ESA) mission
{\it Gaia} (\url{https://www.cosmos.esa.int/gaia}), processed by the {\it Gaia}
Data Processing and Analysis Consortium (DPAC,
\url{https://www.cosmos.esa.int/web/gaia/dpac/consortium}). Funding for the DPAC
has been provided by national institutions, in particular the institutions
participating in the {\it Gaia} Multilateral Agreement. All member stars of the cluster in this work will be released online.
Software: Astropy \citep{2013A&A...558A..33A,2018AJ....156..123A}, Mistree \citep{2019JOSS....4.1721N},
galpy \citep{2015ApJS..216...29B}, Scipy \citep{2020NatMe..17..261V}, TOPCAT \citep{tayl05}.

\end{acknowledgements}


\begin{thebibliography}{}
    \bibitem[Allison et al.(2009)]{2009MNRAS.395.1449A} Allison, R.~J., Goodwin, S.~P., Parker, R.~J., et al.\ 2009, \mnras, 395, 1449. doi:10.1111/j.1365-2966.2009.14508.x
    \bibitem[Astropy Collaboration et al.(2013)]{2013A&A...558A..33A} Astropy Collaboration, Robitaille, T.~P., Tollerud, E.~J., et al.\ 2013, \aap, 558, A33. doi:10.1051/0004-6361/201322068
    \bibitem[Astropy Collaboration et al.(2018)]{2018AJ....156..123A} Astropy Collaboration, Price-Whelan, A.~M., Sip{\H{o}}cz, B.~M., et al.\ 2018, \aj, 156, 123. doi:10.3847/1538-3881/aabc4f
    \bibitem[Bai et al.(2022)]{2022RAA....22e5022B} Bai, L., Zhong, J., Chen, L., et al.\ 2022, Research in Astronomy and Astrophysics, 22, 055022. doi:10.1088/1674-4527/ac60d2
    \bibitem[Bailer-Jones(2015)]{2015PASP..127..994B} Bailer-Jones, C.~A.~L.\ 2015, \pasp, 127, 994. doi:10.1086/683116
    \bibitem[Bisht et al.(2019)]{2019MNRAS.482.1471B} Bisht, D., Yadav, R.~K.~S., Ganesh, S., et al.\ 2019, \mnras, 482, 1471. doi:10.1093/mnras/sty2781
    \bibitem[Bonatto \& Bica(2009)]{2009MNRAS.397.1915B} Bonatto, C. \& Bica, E.\ 2009, \mnras, 397, 1915. doi:10.1111/j.1365-2966.2009.14877.x
    \bibitem[Bovy(2015)]{2015ApJS..216...29B} Bovy, J.\ 2015, \apjs, 216, 29. doi:10.1088/0067-0049/216/2/29
    \bibitem[Bressan et al.(2012)]{2012MNRAS.427..127B} Bressan, A., Marigo, P., Girardi, L., et al.\ 2012, \mnras, 427, 127. doi:10.1111/j.1365-2966.2012.21948.x
    \bibitem[Cantat-Gaudin \& Anders(2020)]{2020A&A...633A..99C} Cantat-Gaudin, T. \& Anders, F.\ 2020, \aap, 633, A99. doi:10.1051/0004-6361/201936691
    \bibitem[Carrera et al.(2019)]{2019A&A...627A.119C} Carrera, R., Pasquato, M., Vallenari, A., et al.\ 2019, \aap, 627, A119. doi:10.1051/0004-6361/201935599
    \bibitem[Chen et al.(2014)]{2014MNRAS.444.2525C} Chen, Y., Girardi, L., Bressan, A., et al.\ 2014, \mnras, 444, 2525. doi:10.1093/mnras/stu1605
    \bibitem[Darma et al.(2021)]{2021MNRAS.506.4603D} Darma, R., Arifyanto, M.~I., \& Kouwenhoven, M.~B.~N.\ 2021, \mnras, 506, 4603. doi:10.1093/mnras/stab1931
    \bibitem[de La Fuente Marcos \& de La Fuente Marcos(2009)]{2009A&A...500L..13D} de La Fuente Marcos, R. \& de La Fuente Marcos, C.\ 2009, \aap, 500, L13. doi:10.1051/0004-6361/200912297
    \bibitem[de la Fuente Marcos \& de la Fuente Marcos(2010)]{2010ApJ...719..104D} de la Fuente Marcos, R. \& de la Fuente Marcos, C.\ 2010, \apj, 719, 104. doi:10.1088/0004-637X/719/1/104
    \bibitem[Dias et al.(2021)]{2021MNRAS.504..356D} Dias, W.~S., Monteiro, H., Moitinho, A., et al.\ 2021, \mnras, 504, 356. doi:10.1093/mnras/stab770
    \bibitem[Dinnbier et al.(2022)]{2022A&A...660A..61D} Dinnbier, F., Kroupa, P., \& Anderson, R.~I.\ 2022, \aap, 660, A61. doi:10.1051/0004-6361/202142082
    \bibitem[Ester et al.(1996)]{Ester1996} Ester, M., Kriegel, H.-P., Sander, J., \& Xu, X.\ 1996, in Proceedings of the Second International Conference on Knowledge Discovery and Data Mining, KDD’96 (AAAI Press), 226
    \bibitem[Fabricius et al.(2021)]{2021A&A...649A...5F} Fabricius, C., Luri, X., Arenou, F., et al.\ 2021, \aap, 649, A5. doi:10.1051/0004-6361/202039834
    \bibitem[Gaia Collaboration et al.(2016)]{2016A&A...595A...1G} Gaia Collaboration, Prusti, T., de Bruijne, J.~H.~J., et al.\ 2016, \aap, 595, A1. doi:10.1051/0004-6361/201629272
    \bibitem[Gaia Collaboration et al.(2018)]{2018A&A...616A...1G} Gaia Collaboration, Brown, A.~G.~A., Vallenari, A., et al.\ 2018, \aap, 616, A1. doi:10.1051/0004-6361/201833051
    \bibitem[Gaia Collaboration et al.(2022)]{2022arXiv220606207G} Gaia Collaboration, Drimmel, R., Romero-Gomez, M., et al.\ 2022, arXiv:2206.06207
    \bibitem[Hao et al.(2022)]{2022A&A...660A...4H} Hao, C.~J., Xu, Y., Wu, Z.~Y., et al.\ 2022, \aap, 660, A4. doi:10.1051/0004-6361/202243091
    \bibitem[Hottier et al.(2021)]{2021A&A...655A..68H} Hottier, C., Babusiaux, C., \& Arenou, F.\ 2021, \aap, 655, A68. doi:10.1051/0004-6361/202140475
    \bibitem[Katz et al.(2022)]{2022arXiv220605902K} Katz, D., Sartoretti, P., Guerrier, A., et al.\ 2022, arXiv:2206.05902
    \bibitem[Krone-Martins \& Moitinho(2014)]{2014A&A...561A..57K} Krone-Martins, A. \& Moitinho, A.\ 2014, \aap, 561, A57. doi:10.1051/0004-6361/201321143
    \bibitem[Krumholz et al.(2019)]{2019ARA&A..57..227K} Krumholz, M.~R., McKee, C.~F., \& Bland-Hawthorn, J.\ 2019, \araa, 57, 227. doi:10.1146/annurev-astro-091918-104430
    \bibitem[Kroupa(2001)]{2001MNRAS.322..231K} Kroupa, P.\ 2001, \mnras, 322, 231. doi:10.1046/j.1365-8711.2001.04022.x
    \bibitem[Lada \& Lada(2003)]{2003ARA&A..41...57L} Lada, C.~J. \& Lada, E.~A.\ 2003, \araa, 41, 57. doi:10.1146/annurev.astro.41.011802.094844
    \bibitem[Lindegren et al.(2018)]{2018A&A...616A...2L} Lindegren, L., Hern{\'a}ndez, J., Bombrun, A., et al.\ 2018, \aap, 616, A2.doi:10.1051/0004-6361/201832727
    \bibitem[Lindegren et al.(2021)]{2021A&A...649A...2L} Lindegren, L., Klioner, S.~A., Hern{\'a}ndez, J., et al.\ 2021, \aap, 649, A2. doi:10.1051/0004-6361/202039709
    \bibitem[Liu \& Pang(2019)]{2019ApJS..245...32L} Liu, L. \& Pang, X.\ 2019, \apjs, 245, 32. doi:10.3847/1538-4365/ab530a
    \bibitem[Li et al.(2020)]{2020ApJ...901...49L} Li, L., Shao, Z., Li, Z.-Z., et al.\ 2020, \apj, 901, 49. doi:10.3847/1538-4357/abaef3
    \bibitem[Massi et al.(2019)]{2019A&A...628A.110M} Massi, F., Weiss, A., Elia, D., et al.\ 2019, \aap, 628, A110. doi:10.1051/0004-6361/201935047
    \bibitem[Murphy \& May(1991)]{1991A&A...247..202M} Murphy, D.~C. \& May, J.\ 1991, \aap, 247, 202        
    \bibitem[Naidoo(2019)]{2019JOSS....4.1721N} Naidoo, K.\ 2019, The Journal of Open Source Software, 4, 1721. doi:10.21105/joss.01721
    \bibitem[Olmi et al.(2010)]{2010ApJ...723.1065O} Olmi, L., Angl{\'e}s-Alc{\'a}zar, D., De Luca, M., et al.\ 2010, \apj, 723, 1065. doi:10.1088/0004-637X/723/2/1065
    \bibitem[Pang et al.(2021)]{2021ApJ...912..162P} Pang, X., Li, Y., Yu, Z., et al.\ 2021, \apj, 912, 162. doi:10.3847/1538-4357/abeaac
    \bibitem[Pang et al.(2022)]{2022ApJ...931..156P} Pang, X., Tang, S.-Y., Li, Y., et al.\ 2022, \apj, 931, 156. doi:10.3847/1538-4357/ac674e
    \bibitem[Pera et al.(2021)]{2021A&A...650A.109P} Pera, M.~S., Perren, G.~I., Moitinho, A., et al.\ 2021, \aap, 650, A109. doi:10.1051/0004-6361/202040252
    \bibitem[Pettersson \& Reipurth(1994)]{1994A&AS..104..233P} Pettersson, B. \& Reipurth, B.\ 1994, \aaps, 104, 233
    \bibitem[Piecka \& Paunzen(2021)]{2021A&A...649A..54P} Piecka, M. \& Paunzen, E.\ 2021, \aap, 649, A54. doi:10.1051/0004-6361/202040139
    \bibitem[Prisinzano et al.(2018)]{2018A&A...617A..63P} Prisinzano, L., Damiani, F., Guarcello, M.~G., et al.\ 2018, \aap, 617, A63. doi:10.1051/0004-6361/201833172
    \bibitem[Priyatikanto et al.(2016)]{2016MNRAS.457.1339P} Priyatikanto, R., Kouwenhoven, M.~B.~N., Arifyanto, M.~I., et al.\ 2016, \mnras, 457, 1339. doi:10.1093/mnras/stw060
    \bibitem[Qin et al.(2023)]{2023ApJS..265...12Q} Qin, S., Zhong, J., Tang, T., et al.\ 2023, \apjs, 265, 12. doi:10.3847/1538-4365/acadd6
    \bibitem[Ripley(1976)]{Ripley1976} Ripley, B. D.\ 1976, J. Appl. Probab., 13, 255
    \bibitem[Ripley(1979)]{Ripley1979} Ripley, B. D.\ 1979, J. R. Stat. Soc. Ser. B (Methodol.), 41, 368
    \bibitem[Salpeter(1955)]{1955ApJ...121..161S} Salpeter, E.~E.\ 1955, \apj, 121, 161. doi:10.1086/145971
    \bibitem[Skrutskie et al.(2006)]{2006AJ....131.1163S} Skrutskie, M.~F., Cutri, R.~M., Stiening, R., et al.\ 2006, \aj, 131, 1163. doi:10.1086/498708
    \bibitem[Tarricq et al.(2022)]{2022A&A...659A..59T} Tarricq, Y., Soubiran, C., Casamiquela, L., et al.\ 2022, \aap, 659, A59. doi:10.1051/0004-6361/202142186
    \bibitem[Tang et al.(2019)]{2019ApJ...877...12T} Tang, S.-Y., Pang, X., Yuan, Z., et al.\ 2019, \apj, 877, 12. doi:10.3847/1538-4357/ab13b0
    \bibitem[Taylor(2005)]{tayl05} Taylor, M.~B.\ 2005, Astronomical Data Analysis Software and Systems XIV, 347, 29
    \bibitem[Virtanen et al.(2020)]{2020NatMe..17..261V} Virtanen, P., Gommers, R., Oliphant, T.~E., et al.\ 2020, Nature Methods, 17, 261. doi:10.1038/s41592-019-0686-2
    \bibitem[Ye et al.(2021)]{2021AJ....162..171Y} Ye, X., Zhao, J., Zhang, J., et al.\ 2021, \aj, 162, 171. doi:10.3847/1538-3881/ac1f1f
    \bibitem[Yuan et al.(2018)]{2018ApJ...863...26Y} Yuan, Z., Chang, J., Banerjee, P., et al.\ 2018, \apj, 863, 26. doi:10.3847/1538-4357/aacd0d
    \bibitem[Zhang et al.(2020)]{2020ApJ...889...99Z} Zhang, Y., Tang, S.-Y., Chen, W.~P., et al.\ 2020, \apj, 889, 99. doi:10.3847/1538-4357/ab63d4

\end{thebibliography}

%
%
\end{document}